\documentclass[12pt]{iopart}
\usepackage{iopams}
\usepackage{graphicx}
\usepackage{harvard}

\begin{document}
\title[Pareto optimality in the quantum Hawk-Dove Game]{Entanglement gives rise to Pareto optimality in the generalised quantum Hawk-Dove Game}
\author{E A B Lindsell$^1$\footnote{Present address: 2 Crabtree Cottages, Marlborough, Wiltshire, SN8~3HP, UK.} and K Wiesner$^1$}
\address{$^1$ School of Mathematics, University of Bristol, University Walk, Clifton, Bristol, BS8~1TW, UK}
\eads{\mailto{eablindsell@gmail.com}, \mailto{k.wiesner@bristol.ac.uk}}
\begin{abstract}
We quantise the generalised Hawk-Dove Game. By restricting the strategy space available to the players, we show that every game of this type can be extended into the quantum realm to produce a Pareto optimal evolutionarily stable strategy. This equilibrium replaces the inefficient classical one when the entanglement prepared in the game exceeds a critical threshold value, which we derive analytically.
\end{abstract}
\indent 
\maketitle

\section{Introduction}
\citeasnoun{von} introduced a theory of games in their paper \textit{The Theory of Games and Economic Behaviour}. Although their intention was to build a theory for predicting economic variation, the field of game theory has since found a diverse range of unforeseen applications from political science \cite{politicsapp} to evolutionary biology \cite{bioapp}. Games such as the Prisoner's Dilemma (PD) \cite{poundstone} and the Hawk-Dove Game (HDG) \cite{maynard} have proved successful models in evolutionary biology to explain the growth of selfish and aggressive behaviours within a population of a species respectively.

\citeasnoun{meyer} inspired the extension of this theory by asking, ``What happens if these games are generalised into the quantum realm?'' The field of quantum game theory was consequentially established by \citeasnoun{strat}. They reasoned that quantum games may already be played on a molecular level where quantum mechanics dictates the rules. As game theory involves the communication of information (such as a player communicating their choice of strategy to the game's arbiter), we can think about this information as quantum information (as we live in a quantum world), thus providing a natural link between game theory and quantum information theory. They also highlight how the presence of phenomena such as entanglement and superposition can drastically change the observed dynamics in this new framework. Examples include the original dilemma present in the classical PD being removed under the quantisation of the game when the two players have access to a specific two-parameter strategy space \cite{games}. \citeasnoun{experiment} have since demonstrated the first physical realisation of a quantum game using a nuclear magnetic resonance quantum computer. It is these types of new results that spur the continued interest in the field of quantum game theory.

In this paper we quantise the generalised Hawk-Dove Game (HDG). By considering the entanglement of the initial state we show that a unique threshold of entanglement can be derived for the onset of a Pareto optimal evolutionarily stable strategy (ESS) for \textit{every} game of this type. We start with an introduction to the classical HDG, followed by a quantisation of the generalised HDG following the work of \citeasnoun{strat}.

\section{The Classical Hawk-Dove Game}
This game was originally formulated by \citeasnoun{maynard} to model the evolution of aggressive genes within a species. He imagined a species who meet pairwise to compete over a resource of value, $v$. In this contest an individual has the choice of two actions. It may either be aggressive (hawk), or non-aggressive (dove).

When hawk is played against dove, the dove will flee before risking an injury, leaving the resource, $v$, to the hawk. If both choose dove they display to each other with a small cost, $d$ (effort or time), and one takes the resource. If both choose hawk they attack each other, one sustaining a large injury, $i$, with the other taking the resource, $v$. This leads to the symmetric pay-off matrix shown in table \ref{HDpay-offmatrix}.

\begin{table}[ht]
\centering
\begin{tabular}{c|cc}
&H & D\\
\hline
H & ($\frac{v}{2} - \frac{i}{2}$, $\frac{v}{2} - \frac{i}{2}$) & ($v$, 0) \\
D & (0, $v$) & ($\frac{v}{2}$ - $d$, $\frac{v}{2}$ - $d$)
\end{tabular}
\caption{The pay-off matrix for the Hawk-Dove Game. The expected pay-offs for the ``row player" are the left-hand entries, the expected pay-offs for the ``column player" are the right-hand entries.}
\label{HDpay-offmatrix}
\end{table}
As is standard, and indeed intuitive, we assume the hierarchy of values to be,
\begin{equation}
0<2d<v<i.
\end{equation}
These conditions ensure hawks gain a negative expected pay-off when competing against other hawks, while doves gain a positive expected pay-off when competing against other doves.

The classical game gives rise to a mixed ESS $(p^*, 1-p^*)$ \cite{maynard}. This mixed strategy is to play hawk with probability $p^* = \frac{v+2d}{i+2d}$. This equilibrium is inefficient however, as the average pay-off received per round to an individual in a population with this resident strategy is $\left(\frac{i-v}{i+2d}\right)\left(\frac{v}{2} - d\right)$ which is less than the non-aggressive, cooperative pay-off of $\frac{v}{2} - d$.

\subsection{Quantisation}
A few quantisations of this game have already been examined, including models by \citeasnoun{nawaz} and \citeasnoun{hanauske}. \citeasnoun{nawaz} use a density matrix quantisation approach to examine the conditions required on the initial state for a pure strategy NE to exist, but do not investigate the entanglement of these initial states. \citeasnoun{hanauske} also use the model of \citeasnoun{strat} to show that varying the entanglement in the system can produce non-aggressive ESSs. They only consider three specific HDGs however, and only a few fixed values of entanglement. We consider both generalised entanglement, and the most general HDG.

\begin{figure}[h]
\centering
\includegraphics[width=7.5cm]{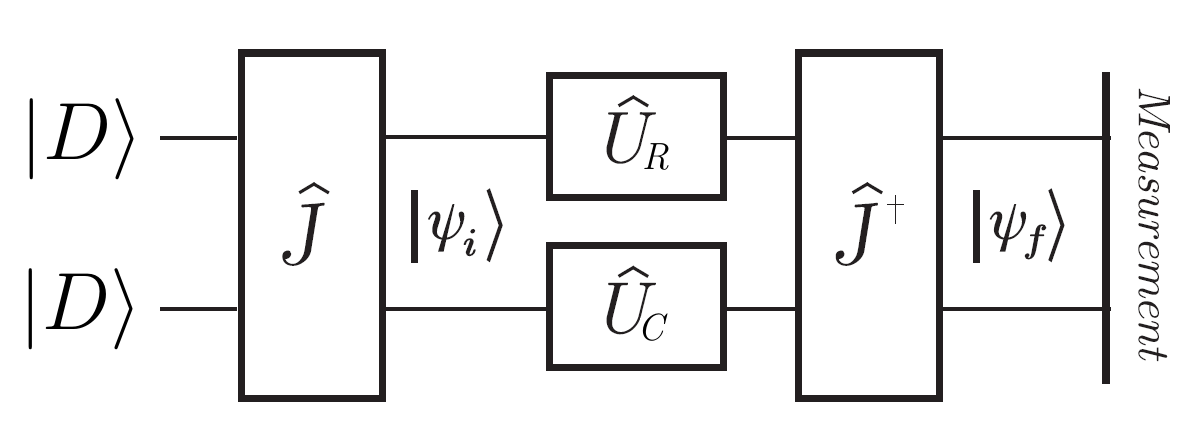}
\caption{The process I use to implement the quantum Hawk-Dove game.}
\label{QHDprocess}
\end{figure}

To quantise this game we use an analogous process to \citeasnoun{strat}, shown in figure \ref{QHDprocess}.
The process is as follows. We start by associating the states $|0\rangle$ and $|1\rangle$ with the strategies dove and hawk respectively. We then assign one qubit in $\mathbb{C}^2$ to each player in the dove state. Hence we describe the state of the game at this point as $|DD\rangle$. This allows us to associate quantum strategies with classical strategies. As a player wanting to use the dove strategy does not want the state of their qubit to change, and a player choosing the hawk strategy wishes to rotate the state of their qubit from $|0\rangle$ to $|1\rangle$, the natural association of operators is,
\begin{equation}
\hat{D} \sim \left(
\begin{array}{c c}
1 & 0 \\
0 & 1
\end{array}
\right),~
\hat{H} \sim \left(
\begin{array}{c c}
0 & 1 \\
-1 & 0
\end{array}
\right).
\label{D&H}
\end{equation}

To generalise the entanglement in this game we use the entangling gate $\hat{J}(\gamma)=\rme^{i\gamma\hat{H}\otimes \hat{H}}$, where we have absorbed the usual power of $1/2$ into gamma for simplicity. This produces the initial state $|\psi_i\rangle=\cos(\gamma)|DD\rangle+i\sin(\gamma)|HH\rangle$.

Here $\gamma \in [0,\pi/4]$ can be viewed as a parameter of entanglement. If $\gamma=0$ the initial state of the game is $|\psi_i\rangle = |DD\rangle$, a product state. If $\gamma=\pi/4,~ |\psi_i\rangle = \frac{1}{\sqrt{2}}\left(|DD\rangle + i|HH\rangle\right)$ which is a maximally entangled state. Thus varying $\gamma$ between $0$ and $\pi/4$ allows us to control the prepared entanglement in the initial state.

Once the qubits have been entangled the players select a unitary operator, $\hat{U}_R \in S_R\subset \mathbb{C}^2 \times \mathbb{C}^2$ ($\hat{U}_C \in S_C\subset\mathbb{C}^2 \times \mathbb{C}^2$) to apply to their qubit. Here $S_R$ ($S_C$) is called the row (column) player's strategy space. After the application of these operators the qubits are disentangled using the gate $\hat{J}^{\dagger}$ and the final state, $|\psi_f\rangle$, is measured. By making a measurement in the computational basis the measured eigenvalues will be either zero or one. A zero corresponds to the measurement of the state $|0\rangle = |D\rangle$, i.e. the corresponding player receives the pay-off for the classical dove strategy. Similarly the eigenvalue one corresponds to the classical hawk strategy.

For simplicity we rewrite the pay-off matrix as,
\begin{table}[ht]
\centering
\begin{tabular}{c|cc}
&H & D\\
\hline
H & ($-a$, $-a$) & ($b$, $0$) \\
D & ($0$, $b$) & ($c$, $c$)
\end{tabular}
\caption{The simplified version of the pay-off matrix for the HDG shown in table \ref{HDpay-offmatrix}, where $a=\frac{i}{2} - \frac{v}{2}$, $b=v$ and $c=\frac{v}{2} - d$, so  $a,b,c>0$.}
\label{HDpay-offmatrixabc}
\end{table}

\subsection{Two-Parameter Strategy Space}
We consider the specific case where each player has access to the two-parameter strategy space, $S$. Thus $S_R=S_C=S$, where $S$ contains all operators of the form,
\begin{equation}
\hat{U}(\theta, \phi) = 
\left(
\begin{array}{cc}
\rme^{i\phi}\cos(\theta) & \sin(\theta) \\
-\sin(\theta) & \rme^{-i\phi}\cos(\theta)
\end{array}
\right)
\label{U2parHD}
\end{equation}
with $0\leq \theta \leq \pi/2$ and $0\leq \phi \leq \pi/2$. Thus the row (column) player's choice of strategy is characterised by their choice of $\theta_R$ and $\phi_R$ ($\theta_C$ and $\phi_C$).

This strategy space was originally constructed by \citeasnoun{strat} to show the existence of a Pareto optimal NE for a maximally entangled game in the quantum Prisoner's Dilemma. We now extend this to the quantum HDG to show the analogous Pareto optimal ESS exists for all HDGs when the entanglement in the initial state, $|\psi_i\rangle$, exceeds a threshold value which we determine.

It is important to observe that the classical game is necessarily contained within this quantisation. By restricting the players to the strategies characterised by $\phi=0$, a player's choice of $\theta\in [0,\pi/2]$ is a bijection to the choice of classical mixed strategies characterised by $p\in[0,1]$.

We calculate the expected pay-off for the row player for generalised entanglement in this two-parameter strategy space using the equation,
\begin{equation}
\$_R(\theta_R, \phi_R, \theta_C,\phi_C, \gamma) = cP_{DD} + bP_{HD}- aP_{HH}, \label{2parHDpay}
\end{equation}
where ${P}_{\sigma{\sigma}^{'}} ~=~ |\langle\sigma\sigma^{'}|\psi_f\rangle|^2$ is the joint probability of measuring $ \sigma, \sigma^{'} \in \{ D,H \}$.

These equations yield the expected pay-off to the row player as,
\begin{equation}
\fl \eqalign{\$_R(\theta_R,\phi_R, \theta_C, \phi_C, \gamma) =c|\cos(\theta_R)\cos(\theta_C)[\cos(\phi_R +\phi_C) + i\cos(2\gamma)\sin(\phi_R + \phi_C)]|^2 \cr
+b|2\cos(\gamma)\sin(\gamma)\sin(\phi_R)\cos(\theta_R)\sin(\theta_C) -\cos(\phi_C)\sin(\theta_R)\cos(\theta_C)[1+i\cos(2\gamma)]|^2 \cr-a|\sin(\theta_R)\sin(\theta_C)+2\cos(\gamma)\sin(\gamma) \sin(\phi_R+ \phi_C)\cos(\theta_R)\cos(\theta_C)|^2.}
\label{2parpay-offHD}
\end{equation}

As this game is symmetric, the column player's pay-off can be found by interchanging the subscripts ``R" and ``C". Note that these pay-offs depend on the entanglement parameter $\gamma$.

We now show that $(\hat{Q},\hat{Q})$ becomes an ESS iff, $\gamma_c \leq \gamma \leq \pi/4,$ where,
\begin{equation}
\hat{Q} = \hat{U}(0,\pi/2)=\left(
\begin{array}{cc}
i&0\\
0&-i
\end{array}
\right).
\end{equation}
I.e. this strategy $\hat{Q}$ becomes the game's new unique ESS when the entanglement in the game exceeds the critical value, $\gamma_c$, which will be determined.

As this is a symmetric game, we can show $(\hat{Q},\hat{Q})$ is a Nash equilibrium by showing $\hat{Q}$ is in the set of best responses to itself, $B(\hat{Q})$, when the entanglement in the game exceeds $\gamma_c$.

Calculating the pay-off received against the strategy $\hat{Q}$ using equation \ref{2parpay-offHD} and looking for the critical values of $\phi$ and $\theta$ that maximise/minimise this pay-off gives the three strategies $\hat{D}$, $\hat{H}$ and $\hat{Q}$. From equation \ref{2parpay-offHD} the expected pay-offs these strategies yield against $\hat{Q}$ are,
\begin{equation}
\$_R(\hat{D}, \hat{Q}, \gamma) = c-4a\cos^2(\gamma)\sin^2(\gamma),
\end{equation}
\begin{equation}
\$_R(\hat{Q}, \hat{Q}, \gamma) = c,
\end{equation}
\begin{equation}
\$_R(\hat{H}, \hat{Q}, \gamma) = b\cos^2(2\gamma).
\end{equation}
Note that $\hat{Q}$ is a best response to itself iff the pay-off,
\begin{equation}
\$_R(\hat{Q}, \hat{Q}, \gamma) \geq \$_R(\hat{U}_R, \hat{Q}, 
\gamma), ~\forall \hat{U}_R \in S.
\end{equation}
\begin{figure}[t]
\centering
\includegraphics[width=7.5cm]{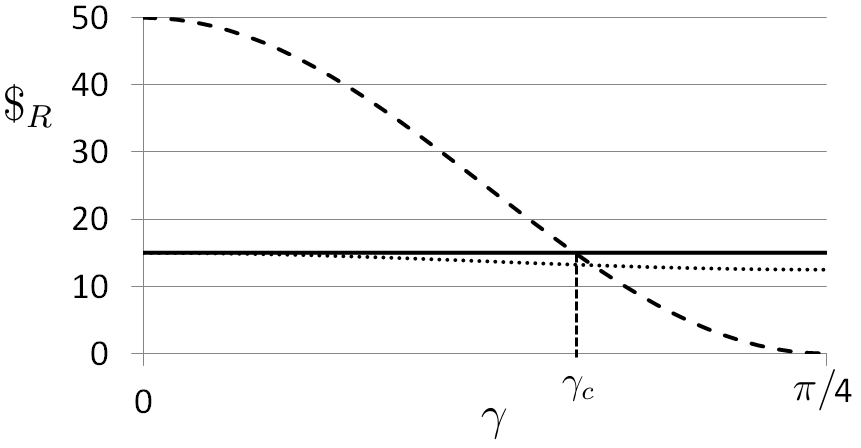}
\caption{The expected pay-off to the row player when playing the strategies $\hat{D} = \hat{U}(0,0)$ (dotted line), $\hat{H} = \hat{U}(\pi/2,0)$ (dashed line) and $\hat{Q} = \hat{U}(0,\pi/2)$ (solid line) against a column player playing $\hat{Q}$, shown as a function of the entanglement parameter $\gamma$. Here we have used the specific values $v$=50, $i$=100 and $d$=10, ensuring $0<2d<v<i$. $\hat{D}$, $\hat{H}$ and $\hat{Q}$ have been chosen because these are the strategies that maximise/minimise a player's pay-off against $\hat{Q}$. This graph shows that $\hat{Q}$ becomes the unique best response to itself when $\gamma_c < \gamma \leq \frac{\pi}{4}$, and is hence an ESS.}
\label{Qvsallgraph}
\end{figure}
We now see that the best response to $\hat{Q}$ depends on the prepared entanglement in the system, as demonstrated in figure \ref{Qvsallgraph}. As $\hat{D}$, $\hat{H}$ and $\hat{Q}$ were found from the turning points of $\theta$ and $\phi$ for all $\gamma$, we need only compare these three strategies. Clearly
\begin{equation}
c \geq c-4a\cos^2(\gamma)\sin^2(\gamma)~ \forall\gamma
\end{equation}
\begin{equation}
\Rightarrow \$_R(\hat{Q}, \hat{Q}, \gamma) \geq \$_R(\hat{D}, \hat{Q}, \gamma)~ \forall\gamma
\end{equation} 
as $c$, $a$, $\cos^2(\gamma)$ and $\sin^2(\gamma)$ are all non-negative. However, we only have 
\begin{equation}
\$_R(\hat{Q}, \hat{Q}, \gamma) \geq \$_R(\hat{H}, \hat{Q}, \gamma)
\end{equation}
iff,
\begin{equation}
c\geq b\cos^2(2\gamma).
\end{equation}
This gives the critical point, $\gamma_c$, when $c= b\cos^2(2\gamma_c)$. So,
\begin{equation}
\gamma_c = \frac{1}{2}\cos^{-1}\left(\sqrt{\frac{c}{b}}\right) = \frac{1}{2}\cos^{-1}\left(\sqrt{\frac{1}{2}-\frac{d}{v}}\right).
\end{equation}

Note that under our original formulation we required $0<2d<v$ to ensure the game remained intuitive, which implies,
\begin{equation}
\frac{\pi}{8} < \gamma_c < \frac{\pi}{4}.
\end{equation}
As $\gamma\in[0,\pi/4]$, this implies $(\hat{Q}, \hat{Q})$ is only an NE iff,
\begin{equation}
\gamma_c \geq \gamma \geq \frac{\pi}{4}.
\end{equation}
Note that $\gamma_c$ always lies strictly within the range of $\gamma$, and hence can always be exceeded. This is shown in figure \ref{Qvsallgraph} where we see that playing $\hat{Q}$ against $\hat{Q}$ is only the best choice of strategy when $\gamma_c \leq \gamma \leq \frac{\pi}{4}$.

By the conditions stated by \citeasnoun{ESS}, $(\hat{Q},\hat{Q})$ is also an ESS in the region $\gamma_c \leq \gamma \leq \frac{\pi}{4}$. Indeed, when $\gamma>\gamma_c$, $\hat{Q}$ becomes the \textit{unique} best response to itself, and is hence an ESS.

As the pay-off received from playing $\hat{Q}$ against $\hat{Q}$ is $c$ (the cooperative, non-aggressive pay-off), this ESS is also Pareto optimal.

\section{Summary and conclusions}
We have shown that for every intuitive HDG as defined by the conditions $0<2d<v<i$, there exists a critical threshold of entanglement, $\gamma_c\in(\frac{\pi}{8},\frac{\pi}{4})$, such that when the entanglement prepared in the initial state exceeds this value, a new Pareto optimal ESS replaces the old, inefficient, classical ESS when the players are restricted to a specific two-parameter strategy space.

In other words, for high enough levels of entanglement, populations of self-interested agents will benefit from being non-aggressive when competing over resources if their strategy space is restricted to the two-parameter set of unitary operators defined in equation \ref{U2parHD}.

Few applications of the quantum HDG are currently known, although \citeasnoun{hanauske} speculate that the quantum HDG may be used in aid of the prevention of the type of economic crisis observed in 2008.

No known applications of this two-parameter strategy space are known. Hence it may prove more relevant for practical application to examine this game when both players are free to apply any quantum operator of their choosing. Under these conditions it would be interesting to investigate whether we still observe an increase in expected pay-off at the game's ESS when compared to the classical result.

This original result gives further insight into the intriguing power of entanglement and the effect of expanding the strategy spaces available to the players in quantum games.

\section*{Bibliography}
\bibliographystyle{jphysicsB} 
\bibliography{refs}

\end{document}